\documentclass{ifacconf}

\usepackage{amsmath}

\usepackage{mathabx}

\usepackage{graphicx,epsf}      
\usepackage{natbib}

\usepackage{pstricks,pst-node}
\usepackage[latin1]{inputenc}
\usepackage{tikz}
\usetikzlibrary{shapes,arrows}

\usepackage{lscape}
\begin{document}
\begin{frontmatter}

\title{A Framework for Plant Topology Extraction Using Process Mining and Alarm Data}

\author[First]{Amir Neshastegaran} 
\author[First]{Ali Norouzifar} 
\author[First]{Iman Izadi} 

\address[First]{Department of Electrical and Computer Engineering, Isfahan University of Technology, Isfahan 84156-83111, Iran \\
(e-mail: a.neshastegaran@ec.iut.ac.ir, ali.noroozi@ec.iut.ac.ir, iman.izadi@cc.iut.ac.ir)}

\begin{abstract}
Industrial plants are prone to faults. To notify the operator of a fault occurrence, alarms are utilized as a basic part of modern computer-controlled plants. However, due to the interconnections of different parts of a plant, a single fault often propagates through the plant and triggers a (sometimes large) number of alarms. A graphical plant topology can help operators, process engineers and maintenance experts find the root cause of a plant upset or discover the propagation path of a fault. In this paper, a method is developed to extract plant topology form alarm data. The method is based on process mining, a collection of concepts and algorithms that model a process (not necessarily an engineering one) based on recorded events. The event based nature of alarm data as well as the chronological order of recorded alarms make them suitable for process mining. The methodology developed in this paper is based on preparing alarm data for process mining and then using the suitable process mining algorithms to extract plant topology. The extracted topology is represented by the familiar Petri net which can be used for root cause analysis and discovering fault propagation paths. The methods to evaluate the extracted topology are also discussed. A case study on the well-known Tennessee Eastman process demonstrates the utility of the proposed method.
\end{abstract}


\end{frontmatter}

\section{Introduction}
Most industrial processes include different parts and devices that interact with each other through material, energy, and information exchange. Due to these interconnections, various parts of a process can affect other parts. 
Hence, there is a correlation as well as causal relationships with temporal lag between different parts of a plant. Discovering and representing these relationships are studied in the literature using topological models (see \cite{man} and the references therein). Plant topology is often illustrated as a directed graph whose nodes and edges represent plant parts and their interactions \citep{phd-th}. These graphical representations of plant topology can be beneficial for various analysis, maintenance, design and management purposes. For instance, since they show the causal relationship between different parts, they can be used for fault diagnosis. One can follow the graph to observe how a certain fault is propagated through the plant Or track backwards to find the root cause of a plant upset. 

Different methods of topology extraction are generally divided into two categories: model-based and data-based \citep{phd-th}. Model-based methods require sufficient process knowledge, understanding different parts of the plant, and analysis of the dynamical equations governing the plant. It also requires studying plant flow diagrams (PFD) and piping and instrumentation diagrams (P\&ID). Plant topology is then extracted using these diagrams and expert's knowledge. Although, many attempts have been made to automate this process, this category of methods is not fully automated yet, and the plant topology is often extracted manually \citep{man}. 

Data-based methods, on the other hand, can be fully automated. In these methods, the interactions are extracted form plant data. The main idea is that, in the plant if X has a causal relationship with Y, then variations in X cause variations in Y after a certain time lag. Therefore, the causal relationships between X and Y can be seen through their trends and described using temporal elements \citep{ct}. Data used here can be process (time trends of the variables) or event data. The former is used more often in plant topology extraction, as it contains more information about the plant behavior. There are a number of methods available in the literature that use process data for topology extraction, including: Granger causality \citep{grg}, transfer entropy \citep{t-ent}, cross correlation \citep{thorn}, partial directed coherence \citep{pdc}, convergent cross mapping \citep{ccm}, and K-nearest neighbors \citep{kn}. Alarm data , however, is less utilized for such purpose and some the proposed methods in the literature are transfer entropy \citep{t-ent-a}, Granger causality \citep{grg-a} and data mining \citep{dm-a}.

In this paper, we propose a method to extract plant topology based on process mining techniques and alarm data. Alarm data is far less dense than continuously generated process data. Hence, it can be stored for much longer periods and is readily available through alarm databases or even flat files. Also, the smaller volume of alarm data compared to process data results in much less computational cost for the processing. On the other hand, the model obtained from alarm data, only covers parts of the plant that contain the alarmed variables. The extracted model, although not complete, is useful for fault analysis as it only includes the parts that are excited during a fault. Therefore, the topology is simpler yet more efficient in fault analysis. Moreover, in the previous studies, the behavior of each process variable or alarm tag is studied in relation with others. Then, these behaviors are combined together to build a model for the topology. This method is somewhat depend on process knowledge -which is only accessible through process experts and operators- in the combination step. However, our proposed method will consider the alarm tags and their relations all together and follow a more systematic procedure which decreases the dependency on process knowledge. The proposed method also enables us to evaluate the extracted topology with respect to the alarm log in addition to common evaluation by process knowledge. 

One point needs to be clarified here to avoid future disambiguation. A process in general (i.e., a series of actions or steps taken in order to achieve a particular end), as in process mining, is somehow different from what a process means in engineering (i.e., a systematic series of mechanized or chemical operations that are performed in order to produce something). Throughout this manuscript, both concepts are used. The particular meaning should be clear form the context.

\section{Process Mining}
In process mining, the objective is to extract process-related information from event data \citep{book}. 
Unlike machine learning in which a process is assumed as a black box and only its inputs and outputs are of importance, in process mining, a comprehensive interpretable model of the process is obtained that can give an inside view of the process \citep{b-box}. For obtaining such model, data should show process transitions and events, as well as flow of information within the process. The data  contains information regarding the beginning and end of an event or transition, and often their sources.  

The data is then analyzed using different process mining techniques to discover a model of the process.
some of the well-known discovery algorithms are alpha miner \citep{alpha}, heuristic miner \citep{heu}, inductive miner (IM) \citep{IM1, IM2}, and evolutionary tree miner (ETM) \citep{ETM}. Alpha miner is a plain algorithm and only useful for simple data. Heuristic miner algorithm focuses on finding frequent  patterns in data. The ETM and IM algorithms discover a model that shows a general overview of the process. 

In addition to mining algorithms, model representation is also important. The most common model representations include: process tree \citep{PT}, Petri net \citep{PN}, business process model and notation (BPMN) \citep{BPMN}, transition system \citep{automaton}, and causal net\citep{c-net}. Each algorithm produces a certain type of model representation. Process tree, Petri net, and BPMN models give a broad view of the process. Transition system model shows different states of the process and their transitions. Causal net model demonstrates the causal relationship between events based on their frequency. In general, the mining algorithm and model representation are closely related. Certain algorithms discover models that have certain representations. So, if a particular model representation is of interest, the consistent algorithm should be used. However, some model representations can be converted into each other.

After a model is discovered, it should be evaluated. This step is known as conformance checking in process mining literature \citep{book}. Four common criteria for conformance checking are fitness, precision, generalization, and simplicity. 
High fitness means the model discovered most of the behavior seen in the event log. High precision indicates that the model doesn't allow for unrelated behavior to the event log, i.e. the fewer this misrepresentation, the better precision. Hence, improving precision avoids model underfitting. Generalization is a measure of how the model discovered based on the given data, can be generalized to the other data of the same process. Improving generalization prevents overfitting. Simplicity shows how simple and comprehensive the model is \citep{book}. 

\section{Process Mining using Alarm Data}
Distributed control systems (DCS) are a fundamental part of process industry nowadays. They manage the data received from a large number of sensors. The sensors measure almost ever process variable (e.g., temperature, pressure, level, and flow). The measurements are time-stamped data, known as process data, and are collected by the DCS and stored in suitable databases. Another set of data that is available in plants is event data, which consists of messages generated by the DCS or other devices when a pre-defined event occurs in the plant. Nearly all event data is related to plant alarms. An alarm is a notification to the operator when an equipment malfunctions, the process deviates from its normal behavior, or due to an abnormal condition that requires attention. For each alarm, a number of messages can be generated by the DCS or other devices, including: alarm activation (ALM), return-to-normal (RTN), and operator acknowledgment (ACK) \citep{ctrl}.

When a fault occurs in some part of the plant, it often propagates through the plant due to process variable interconnections. This results in a number (sometimes very large) of alarms generated in response to a single fault. The alarms usually follow the path of fault propagation and show a temporal pattern related to the underlying fault. Therefore, the alarms related to that fault have a particular time-line. Due to the inherent characteristics of alarm data (they are based on events and have a meaningful time-line) they can be analyzed using process mining algorithms. A model obtained from alarm data using process mining algorithms represents the particular part of the plant topology that is influenced by the fault. In other words, it models the fault propagation path. This topology is obviously not complete (as all the plant is not generally affected by a fault, although this can happen from time to time). But it is simple yet sufficient enough to facilitate fault diagnosis and root cause analysis. This procedure is covered in this paper.

\tikzstyle{decision} = [diamond, draw, fill=blue!20, 
    text width=5em, text badly centered, node distance=3cm, inner sep=0pt, scale=0.7]
\tikzstyle{block} = [rectangle, draw, fill=blue!20, 
    text width=5.5em, text centered, rounded corners, minimum height=2em, scale=0.7]
\tikzstyle{logic} = [circle, draw, fill=blue!20, 
    text width=5em, text centered, inner sep=0pt, scale=0.25]    
    
\tikzstyle{line} = [draw, -latex']
\tikzstyle{cloud} = [draw, ellipse,fill=red!20, node distance=3cm,
    minimum height=2em]
\begin{figure} 

\centering 
 
\begin{tikzpicture}[node distance = 2cm, auto]
    \node [block] (Dataset) {Alarm Dataset};
    \node [block, below of=Dataset, node distance=1.5cm] (TS) {Seperate Transitional States};
    \node [block, below of=TS] (Repetative) {Remove Repetitive Alarm Tags};
    \node [block, below of=Repetative] (Discovery) {Extract Topology};
    \node [block, below of=Discovery, node distance=1.5cm] (Evaluation) {Evaluate Topology};
    \node [decision, below of=Evaluation, node distance=2cm] (Desired) {Desired Topology?};
    \node [block, below of=Desired] (End) {End};

	\node [block, left of=Repetative, node distance=2.5cm] (Insig) {Remove Insignificant Alarm Tags};
	\node [block, left of=Discovery, node distance=2.5cm] (Tune) {Tune Algorithm Parameters};
	\node [logic, left of=Tune, node distance=5.5cm] (OR) {\Huge OR};

	\path [line] (Dataset) -- (TS);
	\path [line] (TS) -- (Repetative);
	\path [line] (Repetative) -- (Discovery);
	\path [line] (Discovery) -- (Evaluation);
	\path [line] (Evaluation) -- (Desired);
	\path [line] (Desired) -- node [scale=0.65]{Yes}(End);
	\path [line] (Desired) -| node [near start][scale=0.65]{No}(OR);
	\path [line] (OR) -- (Tune);
	\path [line] (OR) |- (Insig);
	\path [line] (Insig) -- (Repetative);
	\path [line] (Tune) -- (Discovery); 	
\end{tikzpicture}
\caption{Flowchart of the proposed framework  for plant topology extraction}
\label{f-ch}
\end{figure}
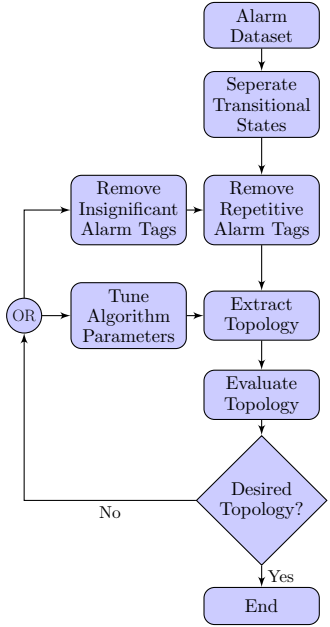

Fig.~\ref{f-ch} illustrates the steps introduced in this paper for process mining based on alarm data. The details of these steps are discussed in the subsequent sections. A case study is then presented on the well-known Tennessee-Eastman process. 

\section{Alarm Data Preparation}
Historical alarm logs contain alarms and other messages generated by the DCS or other plant components. Also operator logs and incident reports often contain related plant faults. In the first step, alarm data should be categorized based on its related plant faults. Assume that a plant has $n$ configured alarm tags. Define  
$ A=\lbrace a_{i}, i=1,\cdots n \rbrace $ as the set of all alarm tags in the plant, where $a_{i}$ is the $i$th configured alarm tag. For a typical plant $n$ may vary from tens to hundreds of thousands. Then all the alarm tags related to each plant fault can be arranged in a sequence in the form of $ \langle e_{1}, e_{2}, \cdots, e_{k}\rangle$ where $e_{j} \in A, j = 1,\cdots,k$. Each sequence is considered as a case. The chronological order of the alarm tags in the sequences is of importance. For example in the aforementioned sequence, alarm tag $e_{1}$ is triggered before alarm tag $e_{2}$ in the historical alarm log. 

In the next step, all the cases related to the same fault, are gathered in one collection and form a dataset. In fact, each dataset is characterized by a specific type of fault. Then, all the alarm sequences of each dataset, are analyzed independently of other faults' datasets. The more cases in the dataset, the better topology can be extracted. 

Datasets can be displayed by dotted chart. As an example, Fig.~\ref{dotted1} illustrates a dataset of 60 cases obtained from the case study which will be discussed later. Each alarm tag in the dotted chart is represented by a unique color. Cases of the dataset are on the vertical axis and time is reflected on the horizontal axis.  Dotted chart is a useful tool for subsequent qualitative and quantitative analysis of alarm data in the datasets.

\subsection{Qualitative Analysis of Alarm Datasets}
After a fault with constant magnitude (e.g., bias in a sensor or power loss in an actuator) occurs in a plant, the plant initially enters a transitional state and then settles in a new operating point or shuts down. The transitional state is defined as a time span which starts with fault occurrence and culminates with a new operating point. The goal here is to extract plant topology in the transitional state, since in this period the fault starts to affect the plant and propagates to other parts. Consequently, it contains the information about plant behavior switching from healthy to faulty states. 
Dotted chart can display the border between the transitional state and the new operating point. As it is observable in Fig.~\ref{dotted1}, in all cases the behavior of alarm tags from the moment of fault occurrence  till a specific point in time (about 21 hours in the figure) are very similar to each other. This period demonstrates the transitional state in which the plant behavior is under the influence of the fault. Later in time, the plant operates according to the new faulty operating point.   

\subsection{Quantitative Analysis of Alarm Dataset}
When a fault occurs in a plant, it may trigger a number of alarms, from a few to a large number (i.e., an alarm flood). This phenomena causes cases in the dataset to encompass a number of alarm tags. Although these cases can be theoretically exploited to extract the plant topology, the obtained topology models might be pointlessly complicated. In fact, a useful topology should be simple enough to be interpretable and meaningful to human operators. To accomplish such topology, the number of alarm tags in the each case must not be very large. However, prior to reducing the number of alarm tags, the expectations of the topology should be determined. This is of high importance both in data preparation and model discovery steps. For instance, the expectation might be to discover the general overview of a plant, finding interactions between sub-processes of a whole plant or detecting the relationships between each alarm tag with others. 

In this paper, topology extraction is discussed which aims to achieve a broad overview of the influenced parts of the plant in the presence of a fault. Thus, the point in which the plant behavior starts to change and how it continues, predominantly matters in this concept. It is expected that the topology is capable of showing fault propagation path. Therefore, reduction of alarm tags in the cases should be done in a way to preserve the necessary information for extracting the expected topology.

\subsection{Dataset preprocessing} 
According to qualitative and quantitative analysis, a dataset needs to be preprocessed. First, it is assumed that the alarm management system is designed and configured appropriately and as a result historical alarm log is clear of unacceptable phenomena such as chattering and stale alarms. If it is not the case, the alarm log should be preprocessed to remove such nuisances \citep{ctrl}. Then, using the dotted chart, alarm data related to transitional states of the plant is maintained and the rest is ignored. 

In the next step, the volume of alarm data related to transitional states should be decreased. The first appearance of each alarm tag in each case, reflects the fact that the fault has reached to the relevant point in the plant. Other appearances of the same alarm tag just emphasize the persistent influence of the fault on the same point of the plant. Hence, to decrease the number of alarm tags in our cases, the first appearance of each alarm tag in every case is preserved and the rest are eliminated. This way, not only the the necessary information about fault propagation is maintained, but also the number alarm tags in cases is reduced to an appropriate proportion. 

\section{Modeling and Process Topology Discovery}
Discovering a plant topology from preprocessed alarm data is the main purpose of this framework. There is a verity of discovering algorithms in process mining discipline, and each of them has a specific model representation. Plant topology is expected to describe behavior of the plant. The model we propose to extract here is used for fault analysis, i.e., determination of fault propagation path and root cause analysis. In process mining, representations such as Petri net, process tree and BPMN provide the desired view of the plant. However, as Petri net is more common and renowned in the  field of control and process engineering, this representation is used in our framework. Hence, a discovery algorithm should output the topology in the form of Petri net, or another representation convertible to Petri net, such as process tree.

\begin{figure}[tb]
\centering 
\includegraphics[scale=0.3]{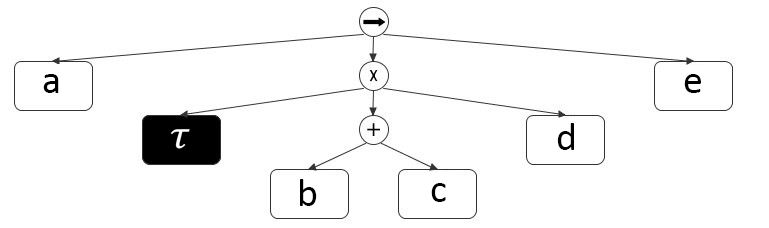}
\caption{Process tree for the sample dataset}
\label{process_tree_sample}
\end{figure}

\begin{figure}[tb]
\centering 
\includegraphics[scale=0.53]{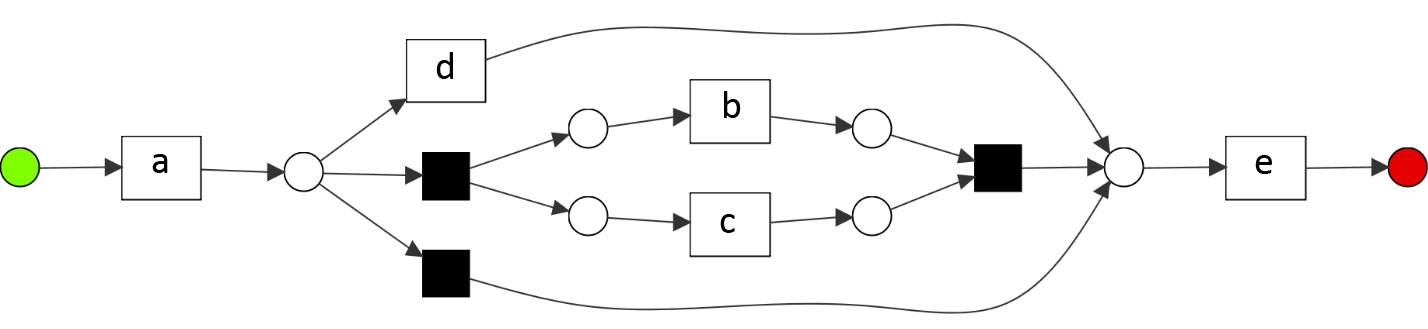}
\caption{Equivalent Petri net for the sample dataset}
\label{petrinet_sample}
\end{figure}

Process tree represents a block-structured hierarchical model. Operator nodes and leaf nodes are the elements of this model. Operator nodes determine how their children relate to each other. Each process tree has an operator set defines what kind of operations are allowable in the model. Generally, these operators are exclusive choice ($\times$), non-exclusive choice($\vee$ or $o$), sequence ($\rightarrow$), parallel execution ($\wedge$ or $+$), and loop execution ($\circlearrowleft$). In some situations, the algorithm requires to execute a silent transition whose execution is not visible in the dataset. We denote this transition with $\tau$. For instance, consider a sample dataset $[\langle a, b, c, e \rangle^{45}, \langle a, c, b, e \rangle^{30}, \langle a, d, e\rangle^{45}, \langle a, e \rangle^{30}]$. The superscript of each sequences indicates its frequency in the dataset. The process tree of the dataset is depicted in Fig.~\ref{process_tree_sample}. Fig.~\ref{petrinet_sample} illustrates the Petri net equivalent of this process tree model.

\subsection{Conformance checking}
After extracting plant topology from the preprocessed alarm dataset, the quality of the extracted model should be evaluated from different perspectives to ensure the model is appropriate and useful. This step is highly depend on the process knowledge. However, the process knowledge is only accessible through process experts and operators and leads to a time consuming procedure. Fortunately, process mining provide us with the concept of conformance checking to evaluate the topology based based on its dataset. In fact, we can benefit from conformance checking to measure the quality of the topology. Therefore, the probable deficiencies can be recognized and resolved prior to referring to the process knowledge evaluation as a time consuming procedure. Conformance checking contains four criteria including \citep{book}: fitness, precision, generalization and simplicity. The extracted topology is then evaluated based on these criteria which are discussed in the following.

\textbf{Fitness}: In process mining, we want to discover a model for our process which can reproduce sequences in our dataset as closely as possible. We use alignment-based approach for computing fitness. For this purpose, we aim to align each sequence in our dataset with the best possible producible sequence in the model.  The highest value of fitness is achieved when all events in the sequence can be aligned synchronously with their corresponding transition in the model. However, often the synchronous alignment is not observed. Therefore, we define a cost function to take into account these non-synchronous moves (such as occurrence of some events without corresponding transition in the model or execution of some transitions in the model without corresponding event in the sequence). Among all possible alignments, one which has the least cost is the optimal alignment. More formal definition of optimal alignment is given in \citep{align}. In short, the following relation is defined to quantify fitness dimension:
\begin{equation*}
F = 1 -  \frac{K(\gamma_{O})}{K(\gamma_{R})}
\end{equation*}
where $K(\gamma_{O})$ is the optimal alignment cost and $K(\gamma_{R})$ is the reference alignment (the worst possible case alignment without any synchronous move) cost to normalize fitness value. With this definition fitness has a value between zero and one.

\textbf{Precision}: This index measures what proportion of possible behaviors of the model are also observed in the dataset. If the model allows for many non-observed behaviors, it means that the model has low precision (i.e., underfitting). Calculation of this index is discussed in \citep{prcs}.

\textbf{Generalization}: This index is proposed to avoid overfitting of the model. It is not desirable to discover a very strict model which can just reproduce existing sequences in the dataset and cannot deal with new sequences from the same process. There are different approaches to calculate this index in process mining literature. For instance, \cite{ETM} used an index in which high generalization is for a model with all parts used frequently. However, in our framework we propose to divide our dataset into training and testing datasets. Then, we measure fitness and precision for both datasets. If similar results are obtained, good generalization is achieved. In this work we randomly choose 2/3 of sequences as training dataset and 1/3 of sequences as testing dataset. 

\textbf{Simplicity}: For operators and process engineers, simple models are more desirable, since they can better follow causalities between variables and predict propagation paths of anomalies. In process mining literature there are quantified indices to measure simplicity such as the one introduced by \cite{ETM}. 

If the discovered plant topology surpasses the predetermined expectations based on the four criteria, the process topology is acceptable. Otherwise, it should be modified. Each discovery algorithm has some adjustable parameters. If for a model some of the indices are not satisfactory, the parameters of the discovery algorithm can be changed in order to improve the model. In addition, if the process topology is not simple enough, we can review the preprocessing step according to \cite{ctrl} and eliminate more alarms unrelated to the specific fault scenario. Afterwards, with this modified dataset, we can discover a simpler process topology.

\section{Model Discovery Algorithms}
An appropriate discovering algorithm should be selected based on industrial plant requirements and expected quality. Among the variety of discovery algorithms developed in recent years for different applications, the Evaluatory Tree Miner (ETM) and the Inductive Miner (IM) algorithms are more compatible with our expectations. ETM is a genetic process mining algorithm and has the ability to factor in different quality dimensions such as fitness, precision, generalization and simplicity during the discovery process. This means we have flexibility to determine the importance of each quality criterion. The IM algorithm has the advantage of discovering a perfectly fit model. This feature is especially beneficial when all the alarm tags in the cases are of critical importance and should be present in the model. In fact, IM is sensitive to infrequent patterns in a dataset (e.g., noisy behavior). However, different versions of IM have been introduced which can better deal with such problems by setting some parameters to filter out infrequent behaviors.

\subsection{The IM Algorithm}
The IM framework in each step tries to find the main cut, which partitions activities (i.e., alarms) into separate non overlapping groups. Then every sequence of dataset is projected on these partitions which leads to a number of sub-sequences relates to each other through the process tree operators ($\times$, $\rightarrow$, $\wedge$, $\circlearrowleft$). These sub-sequences form corresponding sub-datasets. If we organize these sub-sequences according to the operators, the original dataset will be reconstructed. This operator is a node of the process tree. Subsequently, we repeat the same procedure for sub-datasets until we reach a singleton element as a process tree leaf. If the algorithm fails to find new main cut, it instead returns a general model (i.e., a flower model) for this part \cite{IM1}. The output process tree can be converted to a Petri net for further applications.

The IM algorithm is sensitive to noise and infrequent behaviors. Such anomalies prevent the algorithm to find the main cut. In this situation the algorithm discovers a perfectly fitting mode. However, the model lacks precision. This results in non-observed behaviors allowed by the model. To prevent this problem, other algorithms based on IM, have been developed which can better deal with such anomalies. For instance, the IMF algorithm has a filter with adjustable threshold to eliminate infrequent behaviors in the dataset \citep{IM2}. This algorithm, which is used in this paper, first searches for existing cuts. If no cut is found, the log is filtered to remove infrequent behaviors. 

\subsection{The ETM Algorithm}
The ETM algorithm aims to find a model describes the observed behavior according to predetermined priority of quality dimensions. In the first step, the algorithm generates a number of random process trees from the sequences in the dataset. Then, the four quality dimensions (fitness, precision, generalization and simplicity) are calculated for each of the process trees. After that, an overall fitness is calculated as a weighted combination of these four quality dimensions. The wights are assigned to different quality dimensions according to designer's desirable model features. For instance, if precision has more importance in comparison to other dimensions, we can set its weight more than others. In each generation of the algorithm, the best candidates are transferred to the next generation and other candidates change partially or completely based on the overall fitness value. The algorithm continues until the overall fitness reaches the desired value or the algorithm stops in case of a number of unsuccessful attempts or reaching the predetermined maximum number of the generations \citep{ETM}.

\subsection{Comparison of the IM and ETM algorithms}
The IM algorithm discovers a model with perfect fitness but low precision. As it is discussed before, other versions of this algorithm such as IMF improve the model by eliminating noisy behavior. But the priority is still the fitness not the precision. The IM algorithm has better scalability, which means we can use this algorithm for larger and more complex datasets. For large datasets the ETM algorithm has serious restrictions. This algorithm is time consuming and require many computations in each step. On the other hand, compromising different quality dimensions during discovering procedure is a valuable advantage of the ETM algorithm over other algorithms. One can prioritize quality dimensions based on his/her expectations, and the algorithm starts to search for the best model according to the desired values.

Plant topology is useful for a wide variety of applications. If we want to extract a topology with high fitness value or we have large and complex datasets, the IM algorithm is more suitable. However, if we want a precise topology, the ETM algorithm can be a better choice since we can prioritize precision in the algorithm. To fulfill this requirement, the ETM model output has fewer exclusive and non-exclusive choice ($\times$ and $\vee$) and parallel execution ($\wedge$) operators. Instead, there are more sequence ($\rightarrow$) operators. On the other hand, output model of the IM algorithm consists of more exclusive choice ($\times$) and parallel execution ($\wedge$) operators which increase flexibility of model to include all observed behaviors. However, this flexible model achieves less precision as it represents non-observed behavior in the dataset. In summary, if the process tree is converted to a Petri net, we see more sequential transitions in the ETM output model and more parallel transitions in the IM output model.

\section{Case Study: The Tennessee Eastman Process}
The Tennessee Eastman (TE) process is a benchmark in process control for comparison of control, fault detection and monitoring methods. This process has 12 manipulated variables, and 41 process variables. 20 different fault scenarios are defined for this process \citep{TE}. For the case study, we configured an alarm management system with defined High and Low alarms on all the process variables leading to 82 alarm tags. Delay-timers are added to reduce alarm chattering. 

\begin{figure}
	\centering
	\includegraphics[scale=0.15]{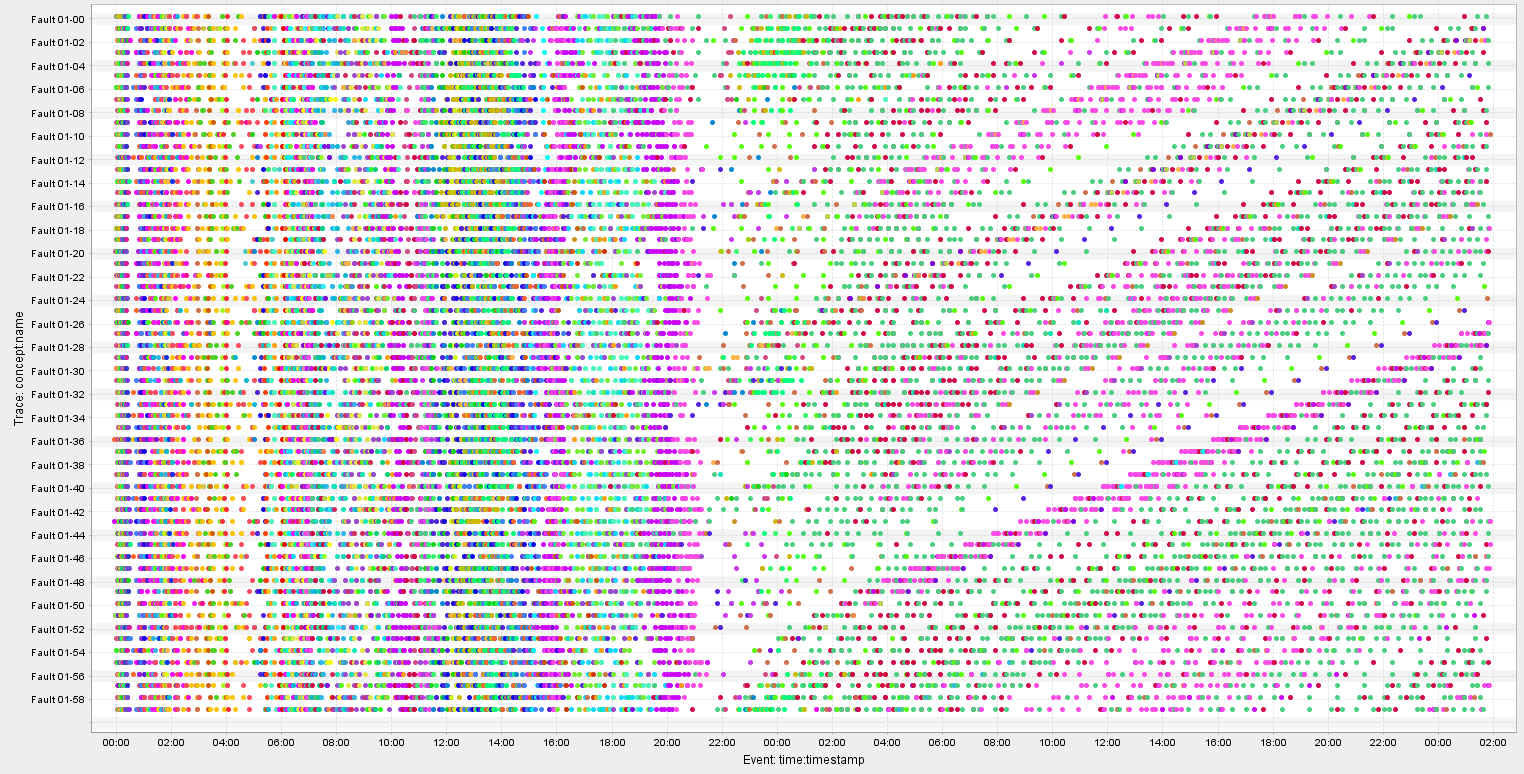}
	\caption{Dotted chart of the dataset (60 cases) related to fault scenario~1 of the TE process}
	\label{dotted1}
\end{figure}

First we need to create a dataset of alarms for the TE process. For this case study, only fault scenario 1 is considered, and the simulation is executed 60 times. Therefore, we have 60 cases in the dataset, all related to fault scenario 1. Fig.~\ref{dotted1} illustrates the dotted chart of this alarm dataset. Due to different random effects in the TE model (noise, disturbances, etc.) the alarms triggered in each case are different. However, a certain pattern can more or less be observed in all the cases.  

\begin{figure}
	\centering
	\includegraphics[scale=0.15]{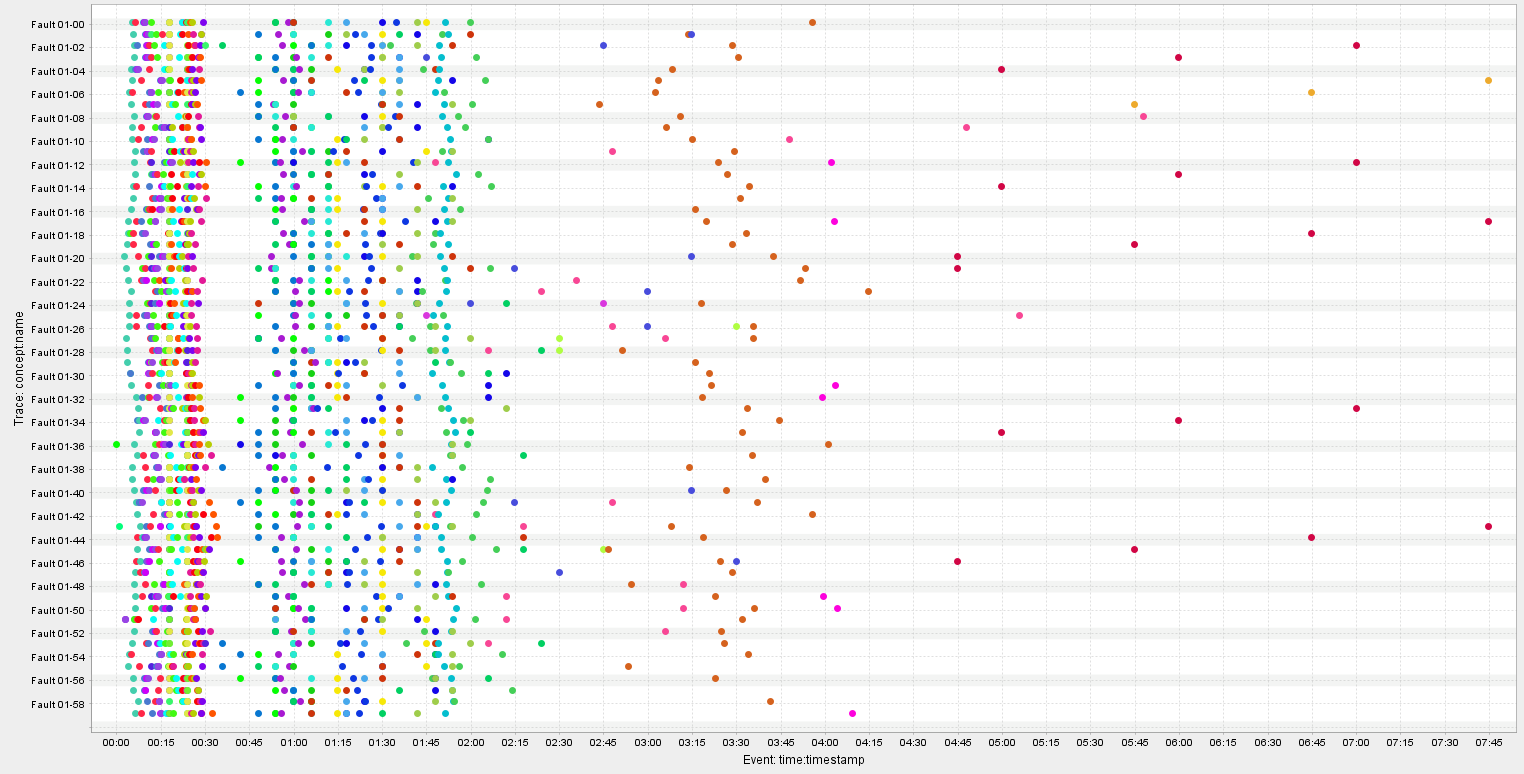}
	\caption{Dotted chart of the preprocessed dataset related to fault scenario~1 of the TE process}
	\label{dotted2}
\end{figure}

As it can be observed from Fig.~\ref{dotted1}, the transient time for this fault lasts about 21 hours, during which a pattern is visible. After this period, the process enters a new state, where the alarms do not relate to the specific fault anymore. Thus, for this analysis, only alarms within the transition period (21 hours) are considered. Also for preprocessing data, repeated and chattering alarms for each case are eliminated. The dotted chart for the preprocessed dataset is illustrated in Fig.~\ref{dotted2}.

\begin{figure}
	\centering
	\includegraphics[scale=0.11]{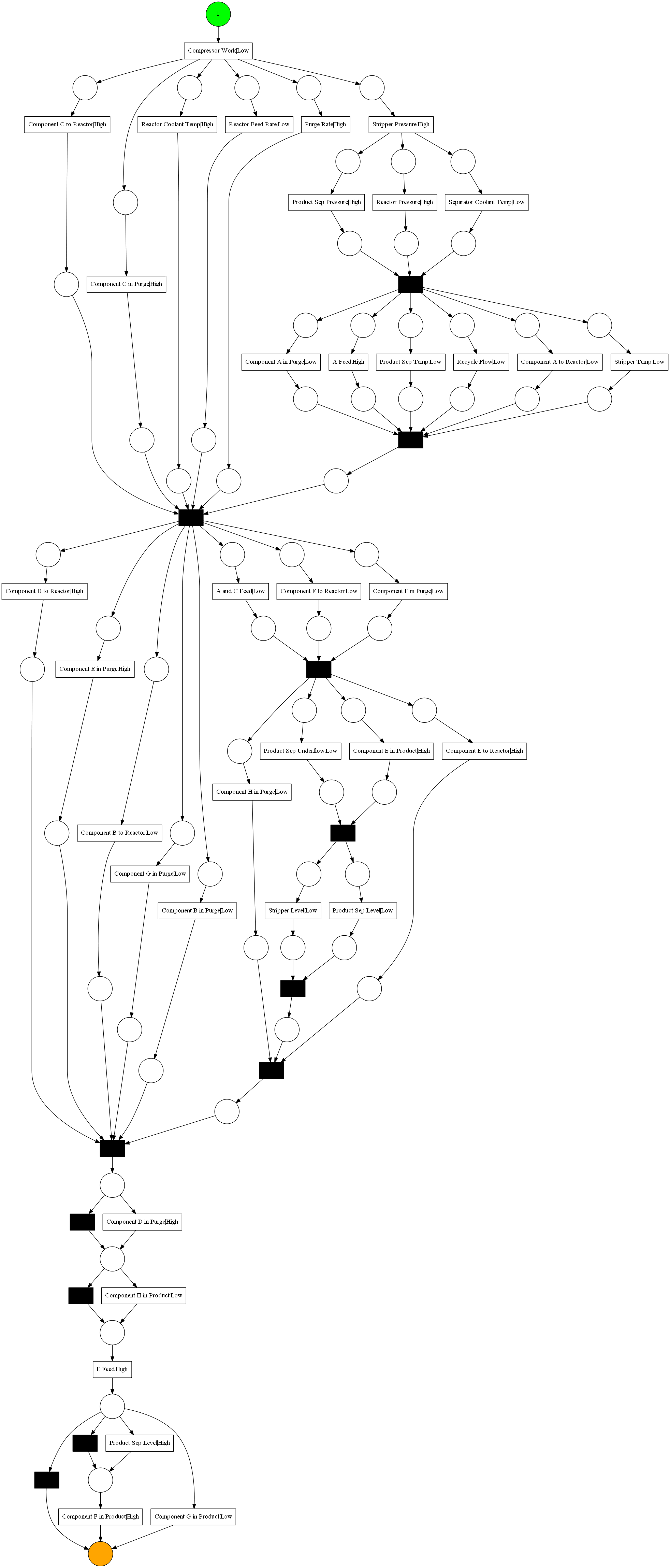}
	\caption{Petri net discovered by the IM algorithm}
	\label{ff1-IM-pn}
\end{figure}

After data preparation, 2/3 of randomly selected cases are separated for training and the rest for testing. These cases are fed to the topology discovery algorithms IM and ETM to extract topology of the process as excited by fault scenario 1. For the IM algorithm the noise threshold is set to 0.3. Fig.~\ref{ff1-IM-pn} shows the Petri net of the process obtained from the IM algorithm. The conformance checking criteria of fitness and precision for this model is also reported in Table~\ref{IM-res}. 

\begin{figure*}[ht]
	\centering
	\includegraphics[scale=0.095]{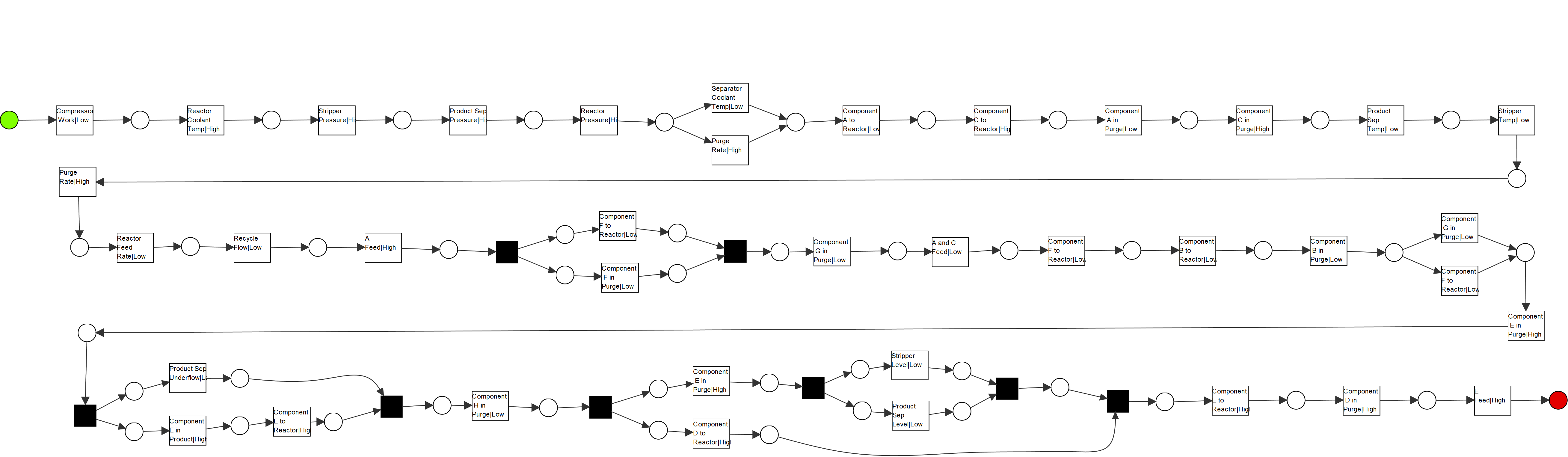}
	\centering
	\caption{Petri net extracted by the ETM algorithm}
	\label{ff1-ETM-pn}
	\centering
\end{figure*}

For the ETM algorithm, the weights for fitness, precision, generalization, and simplicity are selected as 10, 5, 1, and 1, respectively. With these weights, fitness is given the top priority followed by precision. The overall fitness is set to 1 (maximum possible value) and the maximum generation is selected to be 300. Fig.~\ref{ff1-ETM-pn} depicts the Petri net of the process obtained from the ETM algorithm. The conformance checking criteria of fitness and precision for this model is also reported in Table~\ref{ETM-res}.

\begin{table}[b]
\centering
\caption{Evaluation results for the IM algorithm}
\label{IM-res}
\begin{tabular}{|c|c|c|}\hline
 &~Training Dataset~&~Testing Dataset~\\\hline
Fitness&97.3\%&95.8\%\\\hline
Precision&30.9\%&29.6\%\\\hline
\end{tabular}
\end{table}
\begin{table}[b]
\centering
\caption{Evaluation results for the ETM algorithm}
\label{ETM-res}\begin{tabular}{|c|c|c|}\hline
 &~Training Dataset~&~Testing Dataset~\\\hline
Fitness&69.1\%&65.5\%\\\hline
Precision&96.7\%&95.1\%\\\hline
\end{tabular}
\end{table}

As it can be seen from the Petri nets obtained from fault scenario 1, the models are not very complicated and can give a general insight of how fault scenario 1 affects and changes the behavior of the process. For both algorithms the fitness and precision indices obtained form the training dataset are close to that of the testing dataset. This means that the models obtained from both algorithms have acceptable generalization. The main difference between the results obtained from the two algorithms are between the values of their fitness and precision, as well as the general structure of the models obtained. 

Tables~\ref{IM-res} and \ref{ETM-res} show that the IM algorithm yields better performance while the ETM algorithm has better precision. This is absolutely expected based on the properties of the algorithms previously discussed. Also, we expect that the model obtained from the ETM algorithm has more series structure, as in Fig.~\ref{ff1-ETM-pn}. In this structure, the sequence of events is clear which yields a higher precision. But then, in the dataset, there are cases that do not exactly follow this sequence of events, which results in a lower fitness. 

On the other hand, the model obtained based on the IM algorithm has more parallel branches. In this structure, the chronological order of alarms is not exactly determined, i.e., low precision. But then, it can describe more cases in the dataset, i.e., better fitness. So if two cases in the dataset have slightly different order of alarms (which is prevalent in practice), the model can describe both cases. Selection of the preferred algorithm and subsequently the corresponding model, depends on the expectations of the designer.

\section{Conclusion}
In this paper, a method was proposed to extract plant topology from alarm data as an available and highly sparse (compared to process data) source of information. Event based nature of alarm data renders it appropriate for applying process mining techniques. The goal of process mining, too, is obtaining a model of a process using recorded events. Therefore, with redefining the topology extraction problem in process mining context and preparing alarm data, the process mining algorithms can be applied. So, in this paper we used those algorithms to obtain a graphically represented topology of the plant. 

The extracted topology can be evaluated using the four common conformance checking indices: fitness, precision, generalization, and simplicity. The main advantage of these methods over those commonly used in process industry is the use of alarm data instead of process data. This reduces the computational costs and results in a simpler topology. However, the topology obtained here only shows those parts of the plant that are affected by certain faults and contain active alarms. As a result, the topology can be readily used for fault analysis, discovering fault propagation path and root cause analysis. 

From the many known process mining algorithms, the IM and ETM algorithms used in this paper, are more suitable for alarm data. They can generate a model more common in process industry (Petri net). They can also provide a trade-off between fitness and precision (or series and parallel structures) based on the specifications of a particular plant. A case-study on the well-known TE process demonstrated the outcomes of the proposed method. A more detailed study of the TE process and the results that can be achieved based on the obtained topology is under way and will be presented elsewhere.

%


\bibliography{ifacconf}             



\end{document}